\documentclass[aip,jcp,groupedaddress,reprint,twocolumn,a4paper,10pt,longbibliography,amssymb,amsmath]{revtex4-1}

\usepackage{mathptmx}       % selects Times Roman as basic font
\usepackage{helvet}         % selects Helvetica as sans-serif font
\usepackage{courier}        % selects Courier as typewriter font
\usepackage{type1cm}        % activate if the above 3 fonts are
\usepackage{graphicx}        % standard LaTeX graphics tool
\usepackage[bottom]{footmisc}% places footnotes at page bottom
\usepackage{amsmath}  %  e.g., for {equation*}
\usepackage{url}

\begin{document}

\title{Prospects for terahertz imaging the human skin cancer 
with the help of gold-nanoparticles-based terahertz-to-infrared converter}
\author{A.~V. Postnikov}
\email{andrei.postnikov@univ-lorraine.fr}
\affiliation{Universit{\'e} de Lorraine, LCP-A2MC, 1 Bd Arago, F-57078 Metz, France}
\author{K.~A. Moldosanov}
\email{altair1964@yandex.ru} 
\affiliation{Kyrgyz-Russian Slavic University, 44 Kiyevskaya St., Bishkek 720000, Kyrgyzstan}
\author{N.~J. Kairyev}
\affiliation{Kyrgyz-Russian Slavic University, 44 Kiyevskaya St., Bishkek 720000, Kyrgyzstan}
\author{V.~M. Lelevkin}
\affiliation{Kyrgyz-Russian Slavic University, 44 Kiyevskaya St., Bishkek 720000, Kyrgyzstan}

\begin{abstract}
The design is suggested, and possible operation parameters are discussed, of an instrument
to inspect a skin cancer tumour in the terahertz (THz) range, transferring the image into 
the infrared (IR) and making it visible with the help of standard IR camera. The central element 
of the device is the THz-to-IR converter, a Teflon$^{\tiny\textregistered}$ or silicon film matrix
with embedded 8.5~nm diameter gold nanoparticles. 
The use of external THz source for irradiating the biological tissue sample is presumed.
The converter's temporal characteristics enable its performance in a real-time scale. 
The details of design suited for the operation in transmission mode (\emph{in vitro})
or on the human skin in reflection mode (\emph{in vivo}) are specified.
\\*[3mm]
{\sf (To be pubished in the proceedings of the FANEM-2018 workshop -- Minsk, 3-5 June 2018)}
\end{abstract}

\maketitle

\section{Introduction}
\label{sec:intro}
\begin{figure}[b!] % ==================================================================
\centerline{\includegraphics[width=0.49\textwidth]{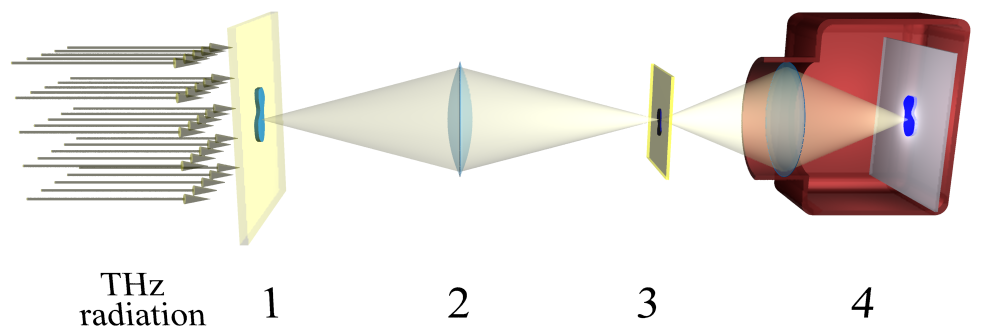}}
\smallskip
\caption{\label{fig:01}%
Suggested setup for \emph{in vitro} studies in the transmission mode. 1 : a tissue sample;
2 : THz objective (high-resistivity float zone silicon, high-density polyethylene
or Teflon$^{\tiny\textregistered}$) with a magnification $M_1$; 3 : THz-to-IR-converter;
4 : highly sensitive IR camera with magnification $M_2$.
Malignant tissue eclipses the bright background.}
\end{figure}

In searching for physical prerequisites to visualize a tumour, one can rely on the THz radiation's 
sensitivity to water, which is one of the most important components of the biological tissue. 
Hu and Nuss \cite{OptLett20-1716} were likely the first to point out the biomedical THz imaging, 
as they noted that the different water content of two different tissues (porcine muscle and fat)
could yield a sufficient contrast. Even earlier, Ross and Gordon \cite{JMicrosc128-7} and
Chen \emph{et al.}\cite{Radiology184-427} have shown that the water content in cancerous tumours
is higher than that in normal tissues. Further on, the water molecules absorb throughout the entire
THz band (0.1 -- 10 THz) \cite{JLaserApp15-192}. 
These properties motivated a development of THz imaging the cancer cells in transmission geometry
(for \emph{in vitro} study of thin, clinically prepared tissue sample, see Fig.~\ref{fig:01}). 
In this approach, the necessary contrast for imaging the tumour and discerning it
from the normal tissue is provided by enhanced water content in cancer cells.

\begin{figure}[b!] % ==================================================================
\centerline{\includegraphics[width=0.49\textwidth]{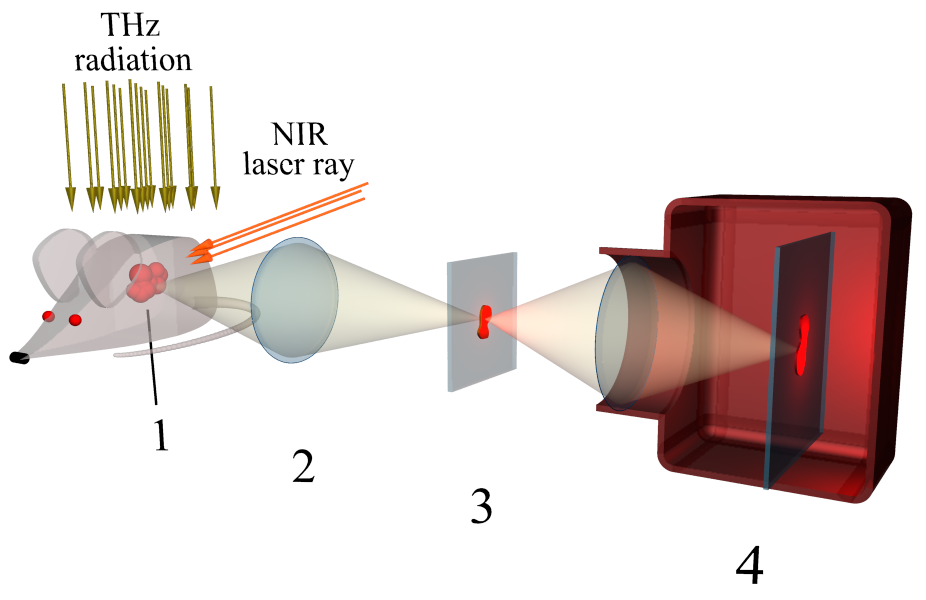}}
\smallskip
\caption{\label{fig:02}%
Suggested setup for \emph{in vivo} imaging in reflection mode.
The near-infrared (NIR) laser serves for excitation of surface plasmons in GNPs inside
a tumour (for heating water in cancer cells). 1 : tumour; 2, 3, 4 : the same as in Fig.~\ref{fig:01}.
Malignant tissue gives a bright signal on a black background.}
\end{figure}

A fact that the water content of cancer cells is higher than that of normal cells, as well as the fact
that the THz waves cannot penetrate moist tissue, favoured the development of another approach
to the THz medical imaging, namely, the reflection geometry (for investigations \emph{in vivo},
Fig.~\ref{fig:02}). Notably when imaging the human skin cancer, the reflection geometry is favoured
over the transmission one, which, in its turn, allows to study the biological tissue samples
of any organ's cancer.

The reflectance of THz radiation, and hence the contrast in imaging the area of cancer, 
is enhanced as the water temperature in cancer cells increases. This observation found its use 
in the THz imaging technique due to works of R{\o}nne \emph{et al.} \cite{JChemPhys107-5319} 
and Son \cite{JAP105-102033},
who studied the power absorption and the index of refraction of water depending on temperature.
Their works prompted progress in optimizing the reflection geometry, \emph{e.g.}, 
by Oh \emph{et al.} \cite{Oh-conf2008,OptExpress17-3469}, who demonstrated 
that the reflected THz signal grows with the water temperature in cancer cells. Further on,
these findings helped to visualise tumour due to its larger contrast, compared with
the background of normal cells in \emph{in vivo}
experiments \cite{Oh-conf2010,OptExpress19-4009,JIRMilliTeraWaves33-74}.

To heat water in cancer cells, the gold nanoparticles (GNPs), like the targeted agents 
in photothermal therapy, are premediatedly delivered into the cancer cells, but not into the normal cells. 
The approach relies on the fact that the targeted agents, antibody-conjugated
gold nanoshells \cite{NanoLett5-709} and solid gold nanospheres \cite{CancerLett239-129},
are accumulated in cancerous tissue more efficiently
than in the normal one. In photothermal therapy, these agents have been demonstrated
to selectively kill cancer cells, leaving the normal cells 
unaffected \cite{NanoLett5-709,CancerLett239-129,JAmChemSoc128-2115}.
Thereupon the tumour is non-invasively treated by irradiating with near-infrared (NIR)
laser beam at $\sim$ 650 -- 1350~nm wavelength; this is the so-called ``therapeutic window''
where light has its maximum depth of penetration into the tissues. Under irradiation, 
the surface plasmons are excited in the GNPs; on dumping out the plasmons, the water is heated
around the nanoparticles in cancer cells. In consequence, the cancer cells start to reflect
the incident THz radiation even more efficiently, and thus can be more readily visualized
by a highly sensitive IR camera (Fig.~\ref{fig:02}).

In works by Woodward \emph{et al.} \cite{SPIE4625-160,PhysMedBiol47-3853}, 
the feasibility of the THz imaging of the body with skin cancer
in reflection geometry have been demonstrated. 

Refs. \onlinecite{JInvestDermatol120-72,BJD151-424,ApplSpectro60-1127}
seem to be the pioneering studies
that revealed a possibility of the THz imaging of the basal cell carcinoma.
It was shown by Wallace and Pickwell \cite{ApplSpectro60-1127,JPhysD39-R301}
that the maximum difference in refractive index between diseased
and healthy tissue occurs at 0.35 -- 0.55~THz. The fact that the maximum difference in absorption
occurs in the vicinity of 0.5~THz strengthens the choice of the above frequency range
as that providing the best imaging contrast when irradiating the biological tissue.

\section{General idea and setup of the terahertz-to-infrared converter}
\label{sec:2} % ====================================================

\begin{table*}[t!] % -----------------------------------------------------------
\caption{\label{tab:01}
Operating frequencies of the THz-to-IR converter and corresponding objective's resolutions.}
\begin{tabular}{r@{.}lr@{.}lr@{.}lr@{.}lr@{.}lc}
\hline
\multicolumn{2}{c}{
\parbox[c]{0.12\textwidth}{\begin{center}Frequency \newline (THz)\end{center}}} &
\multicolumn{2}{c}{
\parbox[c]{0.12\textwidth}{\begin{center}Wavelength \newline ($\mu$m) \end{center}}} &
\multicolumn{2}{c}{
\parbox[c]{0.13\textwidth}{\begin{center}Photon \newline energy \newline (meV) \end{center}}} & 
\multicolumn{2}{c}{
\parbox[c]{0.17\textwidth}{\begin{center}Photon momentum \newline 
($10^{-25}$~g${\cdot}$cm${\cdot}$s$^{-1}$) \end{center}}} & 
\multicolumn{2}{c}{
\parbox[c]{0.20\textwidth}{\begin{center}Objective's \newline resolution ($\mu$m) \newline 
at $l{\sim}100$~mm \end{center}}} &
\parbox[c]{0.20\textwidth}{\begin{center}Transmission $k_1$ \newline of the 0.1~mm thick \newline 
Teflon$^{\tiny \textregistered}$ film (\%) \end{center}} \\
\hline
\hspace*{14pt}\rule[0mm]{0mm}{3mm} 0&38 & \hspace*{10pt}789&5 &  \hspace*{16pt}1&57 &
\hspace*{20pt}0&84  & \hspace*{10pt}${\sim\,}$395&0$^a$ & ${\sim\,}$95 \\
4&2  &  71&3 & 17&4  &  9&3 & ~${\sim\,}$71&3$^b$ & ${\sim\,}$85 \\
\rule[-1mm]{0mm}{2mm}8&7  &  34&5 & 36&0  & 19&2 & ~${\sim\,}$34&5$^b$ & ${\sim\,}$70 \\ 
\hline
\end{tabular}
${^a}A\,{\sim}$200~mm; ${^b}A\,{\sim}$100~mm.
\end{table*} % -----------------------------------------------------------

The availability of commercial IR cameras with temperature sensitivities of $\sim$12 
to 50~mK \cite{Mirage640P-Series,FLIR-A6700sc} 
allows to suggest two schemes of THz imaging a tumour which are shown
in Figs.~\ref{fig:01} and \ref{fig:02}. These imply an usage of the source of the THz radiation
for irradiating the tissue sample \emph{in vitro} or tumour in skin \emph{in vivo}.
In our work \cite{BeilJNano7-983}, we elaborated an
idea that gold nanobars and nanorings irradiated by microwaves could become
THz emitters with photon energies within the full width at half maximum (FWHM) of the longitudinal
acoustic phononic density of states (DOS) of gold ($\approx\,$16 -- 19~meV, 
\emph{i.e.}, 3.9 -- 4.6~THz), with a maximum at $\approx\,$4.2~THz ($h{\nu}{\approx}\,$17.4~meV).
Further on, in Ref.~\onlinecite{Nanotechnology29-285704}
we have shown that gold nanorhombicuboctahedra could be used as emitters of radiation at
``soft'' (two-phonon difference frequency) 0.54~THz
and at ``hard'' (summary frequency) 8.7~THz. The $\simeq\,0.5$~THz
emission is important for the THz imaging of human skin cancer, 
due to maximal contrast it yields between the cancer and the normal tissue, in the context
of findings of Wallace and Pickwell \cite{ApplSpectro60-1127,JPhysD39-R301}.
In the present work, the 0.38~THz radiation will be discussed,
instead of 0.54~THz, in numerical estimates, for the reasons explained below.
The 8.7~THz radiation, due to its shorter wavelength (see Table~\ref{tab:01}), 
might be advantageous for enhancing the spatial resolution of the image.
The combined use of imaging at these two frequencies might be quite promising.

Such gold-nanoobjects-based THz sources are expected to be used in schemes of Figs.~\ref{fig:01}
and \ref{fig:02}, therefore, below, when studying feasibility of the THz-to-IR gold-nanoparticles-based 
converter, we assume that the converter's nanoparticles absorb the THz photons of these three frequencies:
0.38, 4.2, and 8.7~THz. Some corresponding parameters are given in Table~\ref{tab:01}.
The diffraction limit on the objective's resolution ${\Delta}x$ can be estimated as
${\Delta}x{\,\approx\,}\lambda{\cdot}l/A$ ($\lambda$: wavelength, $l$: distance between the objective centre
and the object, hence patient's skin; $A$: the objective's aperture). The estimates show that
these frequencies could yield acceptable resolutions for studying the structure of, say,
pigmentary skin nevi. Further on, in Table~\ref{tab:01}, the transmissions of the 0.1~mm thick
Teflon$^{\tiny \textregistered}$ film at frequencies of interest are given. Below, it will be shown
that this material could be a good choice for the THz-to-IR-converter matrix, 
in accordance with data of Ref.~\onlinecite{THZ-Materials}.

In our work \cite{FANEM2015}, a physical mechanism has been suggested for heating the GNPs
with radiofrequency radiation. At the core of this mechanism are longitudinal 
acoustic vibration modes (LAVMs), which \emph{a priori} could have played a role also in the absorption
of THz photons. In the Subsection~\ref{subsec:absorb-uncertainty}
below, we consider the special case of absorption of THz photons by GNPs 
without involvement of LAVMs, so that the momentum conservation law is fulfilled due to 
uncertainty in momentum of the Fermi electrons of gold.

First, we consider the transmission geometry setup (Fig.~\ref{fig:01}). The idea implies that
an image (in the THz rays) of the tissue sample (position 1 in Fig.~\ref{fig:01})
is projected by the objective (position 2 in Fig.~\ref{fig:01}) onto the two-dimensional
THz-to-IR-converter, which is a matrix, transparent both in THz and IR rays, with embedded GNPs
(position 3 in Fig.~\ref{fig:01}). The GNPs, on irradiation with THz rays, convert the energies
of THz photons into heat, being so the bright spots for subsequent detection by the IR-camera
(position 4 in Fig.~\ref{fig:01}). 

The idea of cancer cells detection is that the latter, rich in water, will strongly absorb
the THz radiation; consequently, the corresponding (eclipsed) areas in the projected image
on the detector plane will generate less heat and appear dark in the resulting IR image,
over the otherwise bright background.

Now we turn to the reflection geometry (Fig.~\ref{fig:02}). In this setup, it is additionally assumed
that the GNPs are delivered to the tumour in advance like the targeted agents in photothermal therapy,
and are irradiated \emph{in situ} with the NIR laser. The laser's radiation penetrates the skin
and the tissue and excites surface plasmons in the therein implanted GNPs. The surface plasmons
heat the tumour, that in its turn enhances reflection of the THz radiation (that is known to be
efficiently scattered by water at elevated temperatures) from it. The THz rays reflected from the tumour
are focused by a lens on the THz-to-IR converter's matrix, creating in this plane a (bright) THz image
of the tumour. Further on, like in the previous case depicted in Fig.~\ref{fig:01}, the GNPs
within the THz-to-IR converter absorb the THz radiation, generate heat and become IR sources,
producing a bright image detectable by the IR camera.

\begin{figure}[b!] % ==================================================================
\includegraphics[width=0.49\textwidth]{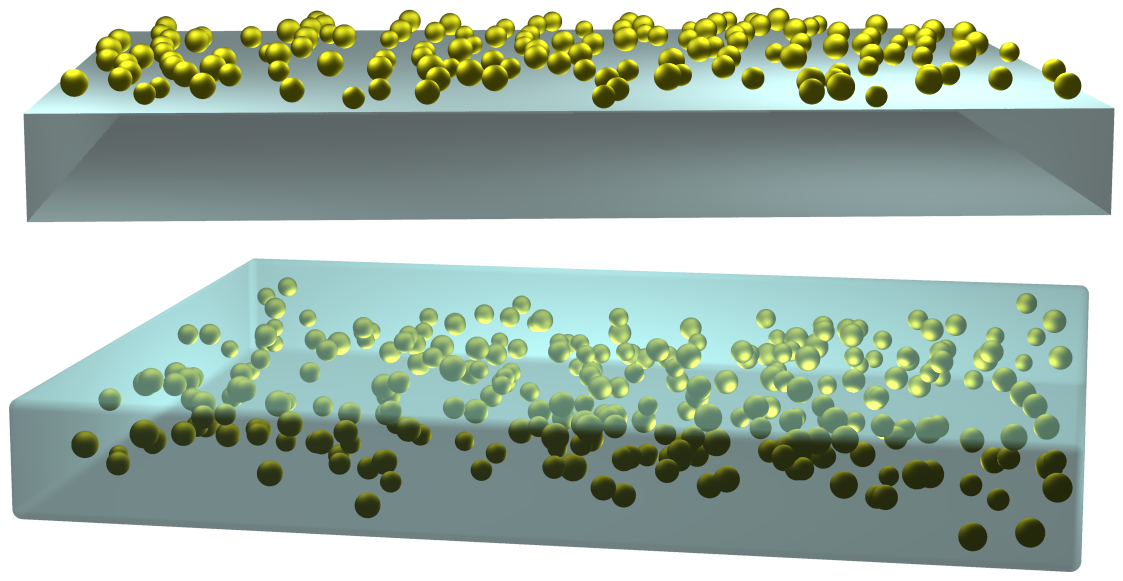}
\smallskip
\caption{\label{fig:03}
Schemes of the THz-to-IR converter, in the form of a substrate transparent
in THz with GNPs deposited onto it (above), or in the form of a matrix
transparent in THz wavelength range with embedded GNPs (below).
}
\end{figure}

Concerning the realisation of the THz-to-IR converter, two schemes shown in Fig.~\ref{fig:03} may come
into discussion. That in the form of a thick film with embedded GNPs (Fig.~\ref{fig:03}, lower scheme)
seems preferable over single-layer deposition (Fig.~\ref{fig:03}, upper scheme), because it allows
to achieve larger ``projected'' density of GNPs per surface unit, avoiding at the same time
to place them too closely. We'll see that these both considerations are important. The spatial resolution
and the depth of focus of the IR camera are the essential parameters to guide the optimal design.

The following considerations help to specify the lens system and sizes of the THz-to-IR converter: 

(1) the objective of the IR camera ensures the close up operation mode with a magnification $M_2=1$.
For infrared cameras, which are kept in mind (Mirage 640 P-Series \cite{Mirage640P-Series}
and FLIR A6700sc MWIR \cite{FLIR-A6700sc}),
this imposes a short distance (${\sim\,}$22 -- 23~mm) between the ``object'' (\emph{i.e.},
the THz-to-IR converter's matrix with GNPs) and the objective's edge, as well as small sizes
(9.6~mm${\,\times\,}$7.7~mm) of the ``object''. The THz-to-IR converter being so small seems
technologically advantageous. At small sizes of the converter's matrix, the distortions caused
by the first lens (the THz objective) would be small, too. The second lens (the IR camera objective)
is simplistically shown in Figs.~\ref{fig:01} and \ref{fig:02}, whereas its real structure
may be much more complex. 

(2) It is convenient to assume the THz objective to be changeable, with different magnifications,
say, $M_1=1;\,0.2;\,5$, whereby the two latter values can be obtained by inverting the same objective. 
Magnification of $M_1=0.2$ would serve a preliminary examination of the patient's skin area,
over the field of view (FOV) of about (9.6~mm${\,\times\,}$7.7~mm)$\,/\,$0.2 = 48~mm${\,\times\,}$38.5~mm.
Operating with the magnification of $M_1=1$, one could inspect a chosen object within the FOV
of 9.6~mm${\,\times\,}$7.7~mm (for example, a pigmentary skin nevi). The magnification $M_1=5$
with its corresponding FOV of (9.6~mm${\,\times\,}$7.7~mm)$\,/\,$5 = 1.92~mm${\,\times\,}$1.54~mm
would enable inspecting details of the pigmentary skin nevi.

The diffraction limits on the objective's resolution ${\Delta}x$ are given in Table~\ref{tab:01}.
At frequency 4.2~THz and $M_2=1$ (in the \emph{close up} operation mode) and pixel size
$d$=15~$\mu$m this means that the linear ``uncertainty'' of the image is covered, on the average,
by ($\sim\,$71.3~$\mu$m)$\,/\,$($\sim\,$15~$\mu$m)$\,\approx\,$5 pixels of the IR camera detector,
that seems acceptable to resolve a meaningfully detailed pattern. At frequencies 0.38 and 8.7~THz
and $M_2=1$ the linear ``uncertainties'' would be covered, on the average,
by $\approx\,$26 and 3 pixels, respectively.

The lenses with desirable parameters (focal distance, diameter) for the THz range,
made out of various materials, can be selected from the lists of commercially available
products -- see, \emph{e.g.}, Ref.~\onlinecite{THZ-Lenses}.

\section{Energy and momentum conservation relations within the photon -- electron -- phonon system}

\subsection{Absorption of THz photons by GNPs 
helped by uncertainty of the Fermi electrons' momenta}
\label{subsec:absorb-uncertainty} % (2.1) ------------------------------------------
\begin{figure}[b!] % ==================================================================
\centerline{\includegraphics[width=0.49\textwidth]{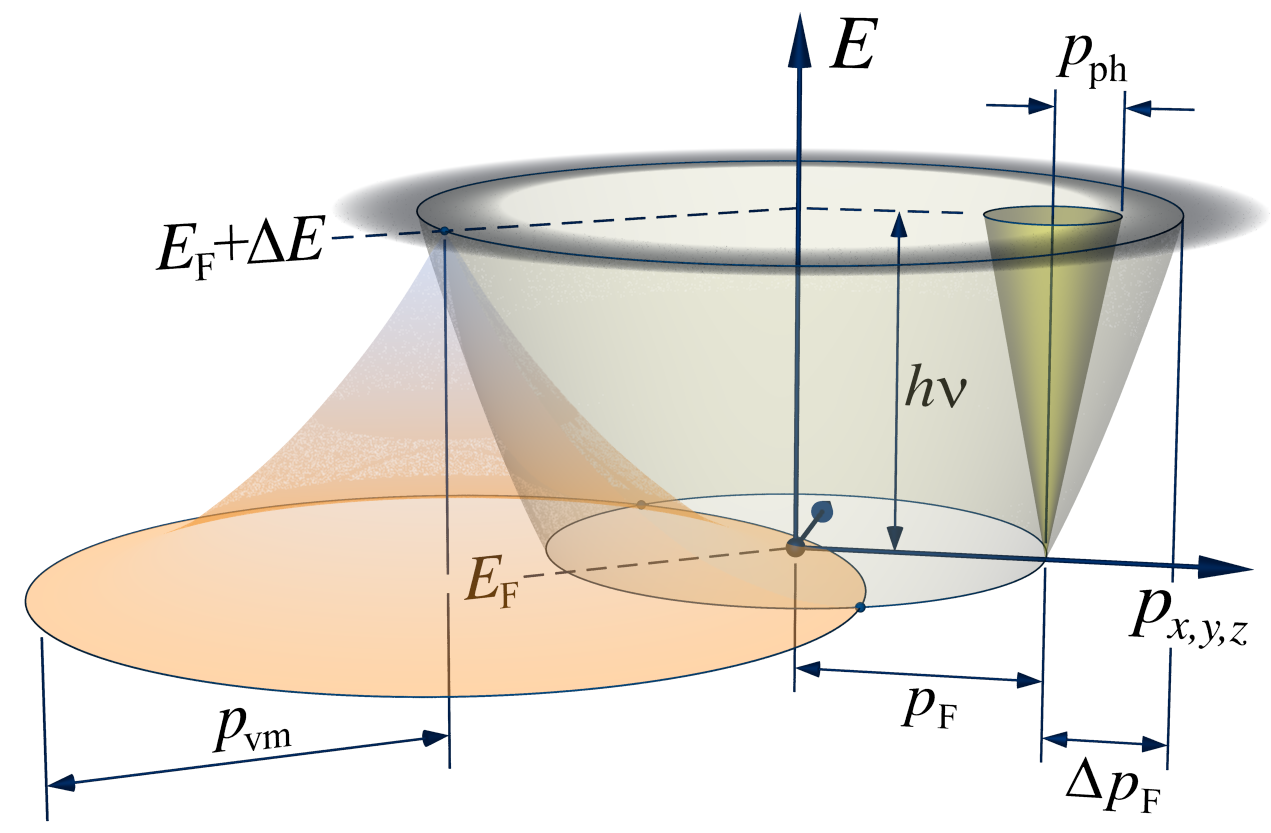}}
\smallskip
\caption{\label{fig:04}
Scheme of absorption of THz photon by the Fermi electron (right part of the figure)
and the latter's subsequent relaxation via releasing a phonon (left part of the figure).
The dispersion law of an electron is depicted by ascending paraboloid, that of the photon --
by narrow ascending cone, that of longitudinal phonon -- by descending bell mouth.
Dispersions are shown over two-dimensional momentum plane. The crossing points 
of the electron and phonon dispersion surfaces at $E=E_{\rm F}$, marked by dots,
indicate relaxation points for the excited electron.
See text for details.}
\end{figure}

We come now to discussion of the physical background of the THz to IR conversion. Let us consider
a situation in GNP, when the energy interval between the Fermi level and
a nearby accessible vacant electron energy level equals the energy of the THz photon, $h\nu$. 
Fig.~\ref{fig:04} depicts a case of absorption of a THz photon by Fermi electron
without involvement of the LAVM; the subsequent relaxation of the excited electron releases
a longitudinal phonon.
The surface of revolution around the vertical (energy) axis in Fig.~\ref{fig:04}
schematically shows the energy dispersion of free electrons (as function of just two momentum
coordinates), delimited at the bottom by Fermi momentum / energy and at the top -- by
a discrete energy level accessible following an excitation. The smearing in the upper momentum plane
indicates the uncertainty of the electron momentum due to spacial confinement (see below), that helps 
to match the momentum conservation condition involving a photon. The dispersion of a THz photon
is shown by a narrow cone, starting from some momentum / energy value of a Fermi electron. 
In the follow-up of an excitation to the energy $E_{\rm F}\!+\!{\Delta}E$, the electron can relax releasing
a longitudinal phonon. The phonon dispersion is depicted by a descending (inverted) 
bell mouth, starting from some momentum / energy of the excited electron.
Essential observations are that ($i$) for small enough nanoparticles, the momentum and energy
conservation in the course of an immediate electron excitation by a THz photon can be assured
by the uncertainty relation; ($ii$) following the relaxation of an excited electron via releasing
a phonon, the electron's momentum may undergo a large reorientation (largely retaining
its magnitude). Let us now consider some details more attentively.

First we try to estimate the sizes of gold nanospheres which could be efficiently used
in the THz-to-IR converter.
The nanoparticle diameter $D$ allows to ``tune'' the electron energy separation ${\Delta}E_{\rm el}$,
according to the Kubo formula for the level spacing in a nanoparticle, 
\cite{JPSJ17-975,JPhysColloq38-C2-69} 
(see also Appendix 1 in Ref.~\onlinecite{FANEM2015}). The energy delivered by a THz photon $h{\nu}$ should match 
an integer number of steps, $m_{\rm el}$, in the electron excitation ladder; simultaneously
it should fit an integer number of energy steps $n_{\rm vm}$ for the vibration mode, ${\Delta}E_{\rm vm}$:
\begin{equation}
m_{\rm el}\,{\Delta}E_{\rm el} = h{\nu} = n_{\rm vm}\,{\Delta}E_{\rm vm}\,.
\label{eq:01}
\end{equation}
For simplicity and order-of-magnitude estimates, we assume
for the beginning a linear dispersion law for the longitudinal phonons, supposing that
they are propagating along the linear size for which we take the diameter $D$, 
possessing the nominal sound velocity $v_{\rm L}$, hence ${\Delta}E_{\rm vm}=v_{\rm L}h/D$. 
A reference to the Kubo formula modifies Eq.~(\ref{eq:01}) as follows:
\begin{equation}
m_{\rm el}(4/3)(E_{\rm F}/N)=h{\nu}=n_{\rm vm}{\cdot\,}v_{\rm L}{\cdot}(h/D)\,,
\label{eq:012}
\end{equation}
where $E_{\rm F}$ is the Fermi energy of gold,
$N$ is the number of gold atoms in the nanosphere, that is otherwise the ratio
of the nanosphere's volume $V=(4/3){\pi}(D/2)^3$ to the volume per atom in the gold's fcc lattice,
$V_{\rm at}=a_{\rm Au}^3/4$, hence $N=V/V_{\rm at}\,\approx\,2.09{\,\cdot\,}(D/a_{\rm Au})^3$.
With this, the diameter $D$ is expressed via the $m_{\rm el}$ and $n_{\rm vm}$ parameters
as follows:
\begin{equation}
D\,\approx\,0.798\,a_{\rm Au}^{3/2}\left(\frac{m_{\rm el}}{n_{\rm vm}}\,
\frac{E_{\rm F}}{v_{\rm L}h}\right)^{\!\!1/2}\,,
\label{eq:011}
\end{equation}
or, with the material constants inserted and specifying the units,
\begin{equation}
D\,\approx\,4.23{\cdot}(m_{\rm el}/n_{\rm vm})^{1/2}\;\mbox{nm}.
\label{eq:02}
\end{equation}
It follows from the right parts of Eqs.~(\ref{eq:01}) and (\ref{eq:012}) that
\begin{equation}
\nu = (n_{\rm vm}{\Delta}E_{\rm vm})/h = n_{\rm vm}(v_{\rm L}/D)\,.
\label{eq:021}
\end{equation}
For given $D$ and $\nu$, we can hope that, by force of Eqs.~(\ref{eq:02}) 
and (\ref{eq:021}), certain combinations of $m_{\rm el}$ and $n_{\rm vm}$ would emerge
as ``resonance'' ones. The optimal choice of all related values comes about from
the following considerations:

$(i)$ For securing a sufficient conrast of the image of cancer / normal tissue, 
the frequency $\nu$ ought to be in the range 0.35 -- 0.55~THz.

$(ii)$ The diameter $D$ must exceed $\simeq\,8$~nm,
in order to prevent the GNP's cooling via spontaneous emission of THz photons.  
This observation was guided by our study \cite{arXiv:1808.10607} 
dedicated to a suggested explanation of the size effect 
in the heterogeneous catalysis on GNPs. Nanoparticles with sizes inferior to 7~nm tend to cool down
due to a spontaneous emission of THz photons, hence become useless if the objective is opposite,
to \emph{heat} the GNPs in the THz-to-IR converter by incoming THz radiation. 
On the other side, an increased $D$ would degrade the converter's sensitivity,
since an elevated THz power level will be required to efficiently heat the GNPs.

$(iii)$
$m_{\rm em}$ and $n_{\rm vm}$ must be integer. With respect to the latter condition, 
we acknowledge that it can be imposed only approximately, because the exact quantisation
criteria would be difficult to control under practical variations of GNPs' shapes
and sizes. Still, the smaller the ($m_{\rm em}$, $n_{\rm vm}$) the stronger the importance
of the confinement effects and hence of the resonance conditions in the energy matching.
We note moreover that the estimation of ${\Delta}E_{\rm vm}$ in terms of $v_{\rm L}$
is only approximate. Anyway, accepting the nominal value of the velocity of sound in gold
$v_{\rm L}=3.23{\cdot}10^5$~cm/s, we arrive at the following compromise 
concerning the parameter values: 
$\nu=0.38$~THz, $D\,{\approx}\,8.5$~nm; $m_{\rm el}=4$; $n_{\rm vm}=1$,
that was accepted for the following analysis.

\begin{table*}[t!] % ----------------------------------------------
\caption{\label{tab:02}Parameters of gold nanospheres 
with thermal conductivity $\lambda_{1p}=73.65$~W${\cdot}$m$^{-1}{\cdot}$K$^{-1}$ 
and diameter $D=8.5$~nm suitable for detection of soft and hard THz radiation.
Estimations are done for $m_{\rm el}=4$, $n_{\rm vm}=1$. 
See text for details.}
\smallskip
\begin{tabular}{r@{.}l@{\hspace*{12pt}}r@{.}lr@{.}l r@{.}lr@{.}l@{\hspace*{12pt}}r@{.}l}
\hline\noalign{\vspace*{-2pt}} %\smallskip}
\multicolumn{2}{c}{\parbox[c]{0.12\textwidth}{\begin{center}Frequency \newline 
(THz) \vspace*{-12pt} \end{center}}} &
\multicolumn{4}{c}{Phonon momentum \vspace*{-2mm}} & 
\multicolumn{2}{c}{\parbox[c]{0.19\textwidth}{\begin{center} \hspace*{2mm}Skin \newline depth (nm) 
\vspace*{-12pt}\end{center}}} &
\multicolumn{2}{c}{\parbox[c]{0.18\textwidth}{\begin{center}${\Delta}p_{\rm D}$ \newline 
\rule[0pt]{0pt}{9pt}($10^{-21}\,$g$\cdot$cm$\cdot$s$^{-1}$) \end{center} \vspace*{-14pt} }} &
\multicolumn{2}{c}{\parbox[c]{0.18\textwidth}{\begin{center}${\Delta}p_{\rm F}$ \newline 
\rule[0pt]{0pt}{9pt}($10^{-23}\,$g$\cdot$cm$\cdot$s$^{-1}$) \end{center} \vspace*{-14pt} }} 
\\
\cline{3-6}
\multicolumn{2}{c}{~} & \multicolumn{2}{c}{$(h/a_{\rm Au})$ \rule[-4pt]{0pt}{14pt}}
 & \multicolumn{2}{c}{$(10^{-20}$~g$\cdot$cm$\cdot$s$^{-1})$ } 
\\
\hline
\rule[0pt]{0pt}{8pt}
\hspace*{3mm}0&38 & \hspace*{2mm}0&048 & \hspace*{8mm}0&78 & \hspace*{6mm}121&05 &
\multicolumn{2}{c}{} & %\hspace*{4mm}$\geq$1&24   & 
\hspace*{6mm}1&79 \\*[-4pt]
             4&35 &              0&770 &              12&5 &               35&8 & 
              \multicolumn{2}{c}{\hspace*{-16pt}\raisebox{4pt}{$\geq\,$1.24}} &    41&1  
\\*[-2pt]
%\hspace*{3mm}0&38 & \hspace*{2mm}0&048 & \hspace*{8mm}0&78 & \hspace*{6mm}121&05 &
% \hspace*{4mm}$\geq$1&24   & \hspace*{6mm}1&79 \\
%             4&35 &              0&770 &              12&5 &               35&8 & 
%              $\geq\,$4&31 &    41&1  \\
\noalign{\smallskip}\hline\noalign{\smallskip}
\end{tabular}
\end{table*} % -----------------------------------

Last but not least, the skin depth in gold at frequencies of interest exceeds 
this GNP size considerably (see Table~\ref{tab:02}). One can therefore assume
that the electric field penetrates the volume and is
of the same strength throughout the nanoparticle. Moreover, the GNP size should be smaller
than the mean free path of electrons in gold, which is, at 300~K,
$l_{\rm b}{\approx}13.1$~nm \cite{Zhang-NanoMicroHeatTransfer}, in order to facilitate
the estimation of the thermal conductivity.

In addition to the above arguments based on the energy conservation law, we should
pay attention to issues of the momentum conservation. 
On excitation of a Fermi electron by a photon of energy $h{\nu}$,
the electron momentum (assuming a free-electron dispersion law) gets an increase.
The momentum of the absorbed THz photon, $p_{\rm ph}$,
is much smaller and hence cannot help to overcome the mismatch. However, 
at small enough GNP sizes, the uncertainty
in the electron's momentum might well absorb the value of ${\Delta}p_{\rm F}$.

Namely, the modification (an increase) of the Fermi electron's momentum
 in the course of absorbing a THz photon is ${\Delta}p_{\rm F}\,\approx\,h{\nu}/v_{\rm F}$,
where $v_{\rm F}\,{\approx}1.4{\cdot}10^8$~cm/s is the Fermi velocity of electrons
in gold \cite{AshMerm-book},
hence ${\Delta}p_{\rm F}\,{\approx\,}4.73{\cdot}10^{-35}{\cdot}\nu$~g$\cdot$cm.
This should be compared to the Heisenberg uncertainty of the electron's momentum 
in a confined geometry,
$$
{\Delta}p_D\,\geq\,h/(2\pi D)\,,
$$
whence, with Eq.~(\ref{eq:02})
and the parameter values for gold, we get:
$$
{\Delta}p_D\,{\geq}\,2.49{\cdot}10^{-21}(n_{\rm vm}/m_{\rm el})^{1/2}\;\mbox{g$\cdot$cm$\cdot$s$^{-1}$}\,.
$$
Some numerical estimations are given in Table~\ref{tab:02}. It follows that, for the GNP size
we consider, the condition ${\Delta}p_{\rm F}\,{\leq}\,{\Delta}p_D$ is by far
respected. Therefore, the mismatch of the electron momenum to satisfy the momentum conservation
on absorbing a THz quantum 
with frequency from the range of interest for biomedical applications
would be always ``absorbed'' by the Heisenberg uncertainty relation.

\subsection{GNP heating through release of longitudinal phonons}
\label{subseq:heating_GNP} % (3.2) -------------------------------------------
The relaxation of excited electron via releasing longitudinal phonon(s),
that would result in heating the GNP, is illustrated by a ``descending''
scheme in the left part of Fig.~\ref{fig:04}.
From the \textbf{energy conservation} condition, the 
excitation energy ${\Delta}E=m_{\rm el}{\cdot}{\Delta}E_{\rm el}$
should match the energy of a vibration mode, according to Eq.~(\ref{eq:01}).
Fig.~\ref{fig:04} depicts for simplicity
a single-phonon process. One can moreover imagine two-phonon (or, in principle,
multiphonon) processes, following the electron excitation by a ``hard'' (${\sim}8.7$~THz) photon.

From the side of \textbf{momentum conservation}, it is seen from
Fig.~\ref{fig:04}, in which at least the qualitative relations are roughly respected,
that there is no general problem of momentum mismatch like that existing in the case
of absorption of a photon, when the photon dispersion surface (a narrow cone)
was entirely placed inside the free-electron paraboloid. 
Indeed, in case of releasing a phonon,
the \emph{minimal} momentum mismatch, \emph{i.e.}, the opening of the bell mouth
on the left of Fig.~\ref{fig:04} for the case $n_{\rm vm}=1$,
is $p_{\rm vm}\simeq\,h/D$, hence ${\simeq}\,2\pi$ times larger
than ${\Delta}\!p_D$ listed in Table~\ref{tab:02} 
(that makes $7.8{\cdot}10^{-21}$~g$\cdot$cm$\cdot$s$^{-1}$, for $D=8.5$~nm), 
and by far larger than ${\Delta}p_{\rm F}$. 
The momenta values of ``genuine'' longitudinal phonons having 
frequencies of interest, as determined from experimental 
phonon dispersion along the ${\Gamma}X$ direction \cite{PRB8-3493} 
and its numerical fit given in Ref.~\onlinecite{Nanotechnology29-285704},
are also shown in Table~\ref{tab:02}. As could be expected, for $\nu=0.38$~THz
the ``exact'' (experimental) value perfectly fits that from the simple 
linear dispersion model as estimated above. Therefore
the matching conditions for both energy and momentum, represented by intersections of
respective dispersion-law surfaces,
can typically be easily found. Fig.~\ref{fig:04} shows two such ``hits'', indicated by
thick dots, on the circle of radius $p_{\rm F}$ in the basal plane $E=E_{\rm F}$.
In 3-dimensional reciprocal space, such matches will be placed on continuous lines
of intersection of isoenergetic surfaces describing the electron and phonon dispersions.
Quantum confinement conditions for electrons and phonons will select distinct ``spots'' 
along such lines, which will be however smeared by force of the uncertainty relation
and the particles' irregularity and dispersion of sizes.
As already mentioned above in relation with the energy conservation, it is difficult
to elaborate on specific relations in the most general case; however it seems plausible
that the electron/phonon energy/momentum matches can be, in principle, easily found
and will ``work'' in the relaxation mechanism.

A quite general and not trivial observation concerning the
energy / momentum conservation in the course of electron-phonon relaxation is that
the momentum of the relaxing electron may be scattered, in the process, quite far
along the Fermi surface. This follows from the comparison of typical values
of ${\Delta}p_{\rm F}$ and (its superior by orders of magnitude)
phonon momentum in Table~\ref{tab:02}. 

Summarizing, the heating of GNP may occur thanks to a combination
of two circumstances. 
$(i)$ The absorption of the THz photon is accompanied by an excitation of a Fermi electron,
which is only made possible by uncertainty of the latter's momentum.
$(ii)$ The relaxation of the excited electron brings about release of a phonon (or, several phonons),
in which process the uncertainty in the momentum of the longitudinal phonon (or, the combined uncertainty
in case of several phonons) may also play an auxiliary role as it mitigates the exact energy / momentum
matching conditions, however, this is not qualitatively essential. The channeling 
of initial photon energy into a vibration channel amounts to heating the particle.

\section{Estimations of parameters of the THz-to-IR converter}
\label{seq:param} % (# 4) --------------------------------------------
For further quantitative estimates, we consider a model of matrix-based converter 
(Fig.~\ref{fig:03}, lower scheme), with Teflon$^{\tiny\textregistered}$ and silicon
as possible matrix materials, transparent in both THz and IR wavelength ranges 
(specifically, throughout $\sim\,$1 to 5~$\mu$m in IR), and gold nanospheres embedded therein. 

As was mentioned above, according to the data of Ref.~\onlinecite{THZ-Materials}, 
the good transmission of Teflon$^{\tiny\textregistered}$ film
of 0.1~mm thickness at the THz wavelengths of interest makes this material
promising as a substrate to host GNPs. The transmission within the IR spectral range
(about 80 -- 90\% for IR wavelength of 3 to 5~$\mu$m and about 50\% for 9 to 12~$\mu$m)
is also acceptable. For the infrared camera, the Mirage 640 P-Series \cite{Mirage640P-Series}
(specifically, the version operating in the wavelength range from 3 to 5~$\mu$m) could be a good choice,
the FLIR A6700sc MWIR \cite{FLIR-A6700sc},
operating in the same wavelength range, being a reasonable alternative.  

Using the previously identified optimal GNP diameter, we calculate now the temporal characteristics
and spatial distribution of heat generated within an isolated GNP embedded in the matrix,
as well as the power levels of the THz radiation required to maintain the GNPs in thermal equilibrium
at, or above, the temperature sensitivity threshold of IR camera.

\subsection{Radiation power required to hold GNPs at temperatures
defined by their emissivity factor $\alpha$}
\label{subsubsec:power_alpha} % (4.1) - - - - - - - - - - - - - -
The temperature distribution over, and around, the GNP can be described by the heat conduction equation
in spherical symmetry with the source function $q(r)$ taken into account:
\begin{equation}
\rho\,C \frac{\partial T}{\partial t}=\frac{1}{r^2}\,\frac{\partial}{\partial r}
\left(\!\lambda r^2\frac{\partial T}{\partial r}\!\right)+q(r)\,,
\label{eq:05}
\end{equation}
where $T(t,r)$ is the temperature, $\rho$ the volume density, $C$ the specific heat, 
$\lambda$ the thermal conductivity, $q$ the volume density of the heat source, 
$q = Q/[(4/3){\pi}R_0^3]$ ($Q$ being the power at the heat source); 
$T$ depends on the radius $r$ and time $t$.
 
We assume the thermal parameters to be temperature independent and uniform inside and outside
the particle, as follows: 
\begin{eqnarray}
0{\,\leq\,}r{\,\leq\,}R_0 &:& \lambda\!=\!\lambda_1,\;\; \rho\!=\!\rho_1,\;\;
C\!=\!C_1;\;\; q\!=\!Q/[(4/3){\pi}R_0^3]\,, \nonumber \\
R_0{\,\leq\,}r{\,\leq\,}R &:& \lambda\!=\!\lambda_2,\;\; \rho\!=\!\rho_2,\;\;
C\!=\!C_2;\;\; q\!=\!0,
\label{eq:06}
\end{eqnarray}
where $R_0$ is the nanoparticle's radius ($R_0 = 4.25$~nm, as argued 
in Subsec.~\ref{subsec:absorb-uncertainty})
and $R$ is the radius of the Teflon$^{\tiny\textregistered}$ shell, $R\,\gg\,R_0$.
For practical calculations below, $R=500$~nm was taken as an effective ``infinity'',
in view of much larger factual matrix thickness ($\sim\,$0.1~mm) and the assumption that
individual GNPs are too distant to interfere. The initial and edge conditions were as follows:
\begin{equation}
T(0,r)=T_R\,;\quad\quad
\left.\frac{\partial T(t,r)}{\partial r}\right|_{r=0}\!=0\,;\quad\quad
T(t,R)=T_R\,.
\label{eq:07}
\end{equation}
A solution for Eqs.~(\ref{eq:05}-\ref{eq:07}) was sought for by the numerical method of lines
relative to ${\Delta}T=T-T_R$, where $T_R=300$~K, as was described in detail 
in Ref.~\onlinecite{JNanophoton6-061716}.

For nanoparticles, both their specific heats and thermal conductivities depend 
on the nanoparticle's size (see Table~\ref{tab:03}).
According to Gafner \emph{et al.} \cite{PhysMetMetallogr116-568},
the specific heat of GNPs is just slightly above that of the bulk gold.
We extrapolated the specific heat $C_{1p}$ of the 8.5~nm diameter GNP
from the data of Ref.~\onlinecite{PhysMetMetallogr116-568} and estimated it to be
133.7~J$\cdot$kg$^{-1}{\cdot}$K$^{-1}$.
The difference in the thermal conductivity is much more dramatic. 
In fact, the \emph{electron} thermal conductivity on the nanoscale is much higher than
the \emph{lattice} thermal conductivity \cite{PRB82-075418,ActaPhysSin62-026501},
however lower than that of the bulk metal.
When the GNP's characteristic size (diameter) $D$ is smaller than the mean
free path of electrons in the bulk gold, \emph{i.e.},
$l_{\rm b}\,\approx\,36.1$~nm at 300~K \cite{Zhang-NanoMicroHeatTransfer}, 
the thermal conductivity of the nanoparticle $\lambda_{1{\rm p}}$ can be estimated 
as follows \cite{Zhang-NanoMicroHeatTransfer}:
$\lambda_{1{\rm p}}\,\approx\,(\lambda_{1{\rm b}}/l_{\rm b}){\cdot}D\,\approx\,73.65$~Wm$^{-1}$K$^{-1}$,
where $\lambda_{1{\rm b}}=312.8$~Wm$^{-1}$K$^{-1}$ is the thermal conductivity of bulk gold.

\begin{table}[b!] % -----------------------------------------------------------
\caption{\label{tab:03}Parameters of materials used in numerical solutions of the heat equation.}
\smallskip
\begin{tabular}{cr@{$\,=\,$}lr@{$\,=\,$}lr@{$\,=\,$}l}
\hline
\parbox[c]{0.12\textwidth}{Material} &
\multicolumn{2}{c}{\parbox[c]{0.11\textwidth}{%
\begin{center} Volume density \\*[-1pt] (g$\cdot$cm$^{-3}$) 
\end{center}}} &
\multicolumn{2}{c}{\parbox[c]{0.11\textwidth}{%
\begin{center} Specific heat \\*[-1pt] (J$\cdot$kg$^{-1}{\cdot}$K$^{-1}$) 
\end{center}}} &
\multicolumn{2}{c}{\parbox[c]{0.11\textwidth}{%
\begin{center} 
\vspace*{-8pt}
Thermal\\*[-1pt] conductivity \\*[-1pt] (W$\cdot$m$^{-1}{\cdot}$K$^{-1}$) 
\vspace*{-8pt}
\end{center}}} 
\\
\hline
\parbox[c]{0.12\textwidth}{\begin{center} 
\vspace*{-6pt}
Nanoparticle: \\*[-1pt]
$D=8.5$~nm \\*[-1pt] gold sphere 
\vspace*{-5pt}
\end{center}} & 
\hspace*{4mm}$\rho_1$&19.32 & 
\hspace*{3mm}$C_{1{\rm p}}$&133.7 & 
\hspace*{3mm}$\lambda_{1{\rm p}}$&73.65 \\
\hline
\rule[0mm]{0mm}{3mm}
Matrix: Teflon$^{\tiny\textregistered}$ & $\rho_2$&2.12 & $C_2$&1160 & $\lambda_2$&0.26 \\
\rule[-1mm]{0mm}{1mm}
Matrix: Silicon                         & $\rho_2$&2.33 & $C_2$&712 & $\lambda_2$&159 \\
\noalign{\smallskip}\hline\noalign{\smallskip}
\end{tabular}
\end{table} % -----------------------------------------------------------

\begin{figure}[!t] % ------------------------------
\includegraphics[width=0.49\textwidth]{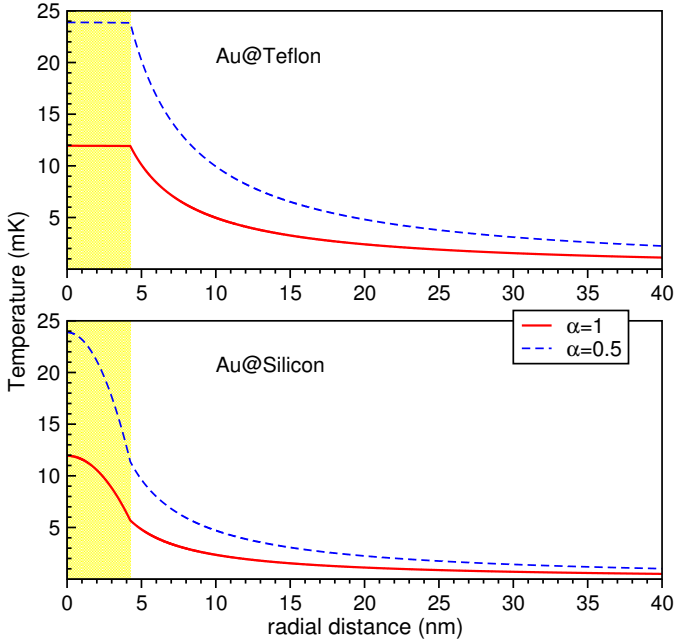}
\caption{\label{fig:05}
Radial distributions of excess temperatures throughout the GNP of radius
$R_0=4.25$~nm (marked by a colour bar) and its embedding by Teflon$^{\tiny\textregistered}$ and silicon
spherical shells, for the values of $Q_{\rm T}$ and $Q_{\rm Si}$ from Table~\ref{tab:04},
corresponding to emissivity factor $\alpha=1$ and $\alpha=0.5$. 
}
\end{figure}

The source term in Eq.~(\ref{eq:05},\ref{eq:06}), stationary
in our assumption, accounts for the power $Q$ delivered to the particle. 
Obviously dependent on the power of the primary THz source, the relevant (threshold) values of $Q$
for our analysis are those which enable a heating, as a steady solution of Eq.~(\ref{eq:05}) -- 
\emph{i.e.}, after an infinite saturation time, -- of the (embedded) GNPs to temperature levels 
detectable by the IR camera. The dynamics of the heating will be discussed below; for the reasons 
elaborated in the next section, we fix such reference threshold value at 12~mK, that is
the reported temperature sensitivity of the \emph{Mirage 640 P-Series IR} thermal imaging camera 
\cite{Mirage640P-Series}.
Individual GNPs, however, may largely vary in their ability to convert absorbed power into heat
(depending on the particle's roughness \emph{etc.}), that can be grasped into a phenomenological 
emissivity factor $\alpha$, varying between 1.0 (ideal $Q$ to ${\Delta}T$ conversion) and almost zero.
Without a clue as for the actual values of $\alpha$, we consider in the following, 
for reference purposes, the values $\alpha=1$ and $\alpha=0.5$, 
the generalisation being straightforward.
The threshold ${\Delta}T_{\alpha}$ values needed to yield visibility by the IR camera
will be scaled by ${\times}\frac{1}{\alpha}$, hence 12 and 24~mK, correspondingly,
for the $\alpha$'s under discussion.
The estimated powers $Q_{\rm T}$ and $Q_{\rm Si}$, defined by GNPs' emissivity factors
$\alpha$, versus the temperature rise ${\Delta}T_m$ for GNP of radius $R_0=4.25$~nm in 
Teflon$^{\tiny\textregistered}$ and silicon spherical shells are given in Table~\ref{tab:04}.
In Fig.~\ref{fig:05}, the radial distributions of temperature around GNPs embedded
in the matrix are shown.

\begin{table}[t] % -------------------------------------
\caption{\label{tab:04}
Parameters of THz-to-IR converter based on GNPs with the raduis $R_0=4.25$~nm,
for two values of emissivity factor $\alpha$.}
\smallskip
\begin{center}
\begin{tabular}{c@{\hspace*{8pt}}c@{\hspace*{8pt}}c@{\hspace*{8pt}}
                c@{\hspace*{8pt}}c@{\hspace*{8pt}}c@{\hspace*{8pt}}c}
\hline\noalign{\smallskip}
 & & \multicolumn{2}{c}{Teflon matrix} && \multicolumn{2}{c}{Silicon matrix} \\
 \cline{3-4} \cline{6-7}
 \rule[0mm]{0mm}{9pt}
$\alpha$ & ${\Delta}T_{\alpha}$~(mK) & 
$Q_{\rm T}$~(nW) & $n_{\rm T}$ & \rule[-2pt]{0pt}{2pt} & $Q_{\rm Si}$~(nW) & $n_{\rm Si}({\times}10^{-3})$ \\
\hline
1   & ~12 & 0.168 & 0.23 & \rule[0pt]{0pt}{9pt} & 49.2 & 0.79 \\
0.5 & ~24 & 0.335 & 0.23 &                      & 98.4 & 0.79 \\
\noalign{\smallskip}\hline\noalign{\smallskip}
\end{tabular}
\end{center}
\end{table}

\subsection{Estimations of THz power threshold sensitivity for
the THz-to-IR converter + IR camera operation}
\label{subsec:power_camera} % (4.2) - - - - - - - - - - - - - - - - - -
On having discussed the ``performance'' of a single embedded GNP for the THz-to-IR conversion,
we turn to an assessment of a realistic device, to be composed of distributed GNPs and
``viewed'' by practically available IR cameras. Let us consider a spot on the converter plate,
which has to be mapped onto the pixel in the IR camera's focal plane array (FPA). 
As was mentioned in the beginning of Section \ref{sec:2}, the IR camera operates in the close up mode,
so that its objective's magnification equals nearly 1. Assuming that $d$ is the pixel size,
the spot area is ${\sim}d^2$. The Stefan -- Boltzmann law yields the energy flux from the area
radiating as the absolutely black body (\emph{i.e.}, the radiant flux surface density) into 
a semisphere (solid angle of $2{\pi}$), in the wavelength range from 0 to $\infty$, as 
$\varepsilon_{bb}={\sigma}T^4$, where ${\sigma}=5.67{\cdot}10^{-8}$~Wm$^{-2}$K$^{-4}$ is
the Stefan -- Boltzmann constant, and $T$ the surface temperature (taken $T=300$~K in the following).
Let us assume that ${\Delta}T_{bb}$ is the nominal temperature sensitivity of the IR camera.
This means that the detectable excess (over the background) in the surface density of the radiant
flux emitted by the reference spot into the solid angle $2{\pi}$ within the whole wavelength range
needs to be at least 
\begin{equation}
{\Delta}\varepsilon_{bb}=4{\sigma}\,T^3{\Delta}T_{bb}\,,
\label{eq:08}
\end{equation}
in order to cause a response of the IR camera.
We assume further on that the surface spot is expanded into a volume element with the same area $d^2$
and thickness $\delta$, inferior to the depth of focus $l_{\rm dof}$ of the IR camera's objective.
We specify that the volume element contains $n$ GNPs with the emissivity factor $\alpha$,
heated by absorbing the THz energy. Those GNPs within $\delta\,{\leq}\,l_{\rm dof}$ heated up to
a temperature of 300~K$+({\Delta}T_{bb}/{\alpha})$ will be perceived by the IR camera. 

The properties of the substrate, in what regards the heat exchange, need to be grasped
by the model to enable realistic estimates of the temporal characteristics and spatial resolution.
Numerical estimations have been done as explained above, assuming that the GNP is placed
into a spherical shell of Teflon$^{\tiny\textregistered}$ (or silicon) of 500~nm radius,
whereby the power of the THz radiation $Q$ heats the GNP in a steady-state mode up to
the temperature of  $({\Delta}T_{bb}/{\alpha})$ above the background --
see a solution of the heat conduction equation (\ref{eq:05}) in subsection \ref{subsubsec:power_alpha},
with the $Q$ values 
for the source term, Eq.~(\ref{eq:06}), as given by Table \ref{tab:04}.
In the steady-state heat transport, the surface density of the radiant flux generated by $n$ GNPs
 element and emitted outwards within the solid angle $4{\pi}$ and the wavelength
range from 0 to $\infty$ is $(k_2{\cdot}n{\cdot}Q)/d^2$, where $k_2$ is the transmission
of the Teflon$^{\tiny\textregistered}$ matrix within the operating wavelength range of the IR camera.
Half of this, \emph{i.e.}, $(k_2{\cdot}n{\cdot}Q)/(2d^2)$, will be emitted in the direction towards
the IR camera. Comparing this latter value against the threshold radiant flux from Eq.~(\ref{eq:08})
specifies the minimal number of GNPs needed in the volume element so that the resulting radiant
flux surface density be sufficient to cause the IR camera's response:
\begin{equation}
n=\frac{8\sigma T^3{\Delta}T_{bb}\,d^2}{k_2 Q}\,,
\;\mbox{or more generally}\;
n=\frac{8\sigma T^3{\Delta}T_{\alpha}\,d^2}{k_2 Q}\,,
\label{eq:09}
\end{equation}
where ${\Delta}T_{\alpha}={\Delta}T_{bb}/\alpha$ (specified in Table \ref{tab:04})
accounts for the emissivity factor $\alpha$ being different from 1.

The $Q$ values to use in Eq.~(\ref{eq:09}), as they follow from calculations described
in subsection \ref{subsubsec:power_alpha}, are listed in Table \ref{tab:04} 
for two trial levels of the emissivity factor $\alpha$ of GNPs. The $Q_{\rm T}$ and $Q_{\rm Si}$ 
are the steady-state THz powers to be delivered to a GNP placed inside the spherical shell made of,
respectively, Teflon$^{\tiny\textregistered}$ or silicon, in order to heat it from temperature
$T_0$ = 300~K to $T_0+{\Delta}T_{bb}/\alpha$. To be specific, the value ${\Delta}T_{bb}$ was chosen
to be 12~mK, the reported temperature sensitivity of the \emph{Mirage 640 P-Series}
infrared thermal imaging camera \cite{Mirage640P-Series}. 

In the model considered, the minimal possible thickness of the film matrix is 1~$\mu$m,
the diameter of the spherical shell covering the GNP. For further assessments,
Teflon$^{\tiny\textregistered}$ film of 0.1~mm thickness would be an acceptable choice
because this is less than the typical depth of focus $l_{\rm dof}$ ($\sim$0.3~mm). Moreover,
the transmission data $k_1$ and $k_2$ in, correspondingly, THz and IR ranges are available for the
Teflon$^{\tiny\textregistered}$ film of namely 0.1~mm thickness \cite{PRB8-3493} --
see the Table~\ref{tab:01}, where the $k_1$ values are given for three frequencies of interest,
whereas $k_2\,\approx\,85$\% within the wavelength range from 3 to 5~$\mu$m, that is,
the operating range of the \emph{Mirage 640 P-Series} infrared camera.

As follows from Table \ref{tab:04}, the values of $n_{\rm T}$ or $n_{\rm Si}$, 
\emph{i.e.}, the ``threshold'' numbers of GNPs within the volume element
that maps onto a pixel of the IR camera, are by far less than 1 in all combinations of the parameters 
considered. In order to ``sensibilize'' each pixel, the number of GNPs within the corresponding
volume element must obviously be, at least, one. An increase in the GNPs concentration
beyond this number ensures each pixel to be successfully addressable.

Assuming $n_{\rm T}$ or $n_{\rm Si} = 1$, the concentration of GNPs in the substrate must be 
$N^{\ast}_{\rm T} = N^{\ast}_{\rm Si} = 1/(d^2{\cdot}\delta)$. Taking the \emph{Mirage 640 P-Series} IR camera
detector as an example, the pixel size is $d=15~{\mu}m$, therefore, for the above justified thickness
of the Teflon$^{\tiny\textregistered}$ layer being 0.1~mm, the GNP concentration is
${\approx}\,4.44{\cdot}10^4$~mm$^{-3}$. This will suffice to make every pixel addressable.

We can further on estimate the power of THz radiation source needed for a Teflon$^{\tiny\textregistered}$
matrix of the 9.6~mm${\times}7.7$~mm size and 0.1~mm thickness to work. The total amount of the GNPs
in the matrix is $N^{\ast}_T{\cdot}9.6~{\rm mm}{\times}7.7~\mbox{mm}{\times}0.1$~mm
${\approx}\,4.44{\cdot}10^4$~mm$^{-3}\,{\cdot}7.39$~mm$^3$ $\approx$ $3.28\,{\cdot}10^5$.
Typically for rough estimates one takes the emissivity factor $\alpha = 0.5$, then the power needed
to heat such GNP by 24~mK is $0.335$~nW, and the required power of the whole THz source
would be \linebreak $3.28{\cdot}10^5{\cdot}\,0.335$~nW $\approx$ 110~$\mu$W. For silicon matrix,
the corresponding values of power are 98.4~nW and 
3.28${\cdot}10^5{\cdot}\,98.4$~nW $\approx$ 32.3~mW, respectively, \emph{i.e.},
larger than those for the Teflon$^{\tiny\textregistered}$ matrix by a factor of $\approx\,$300.

The values 110~$\mu$W and 32.3~mW can be considered as the conventional powers required
for operating the THz-to-IR converters built around the Teflon$^{\tiny\textregistered}$ and 
silicon matrices, respectively, in the transmission mode (Fig. 1). For the reflection mode (Fig. 2), 
these values could classify a sensitivity of the ``THz-to-IR converter + IR camera'' system.

\subsection{Temporal characteristics of the THz-to-IR converter + IR camera system}
\label{subsec:temporal} % (4.3) - - - - - - - - - - - - - - - - - - - 

\begin{figure}[t!]
\includegraphics[width=0.49\textwidth]{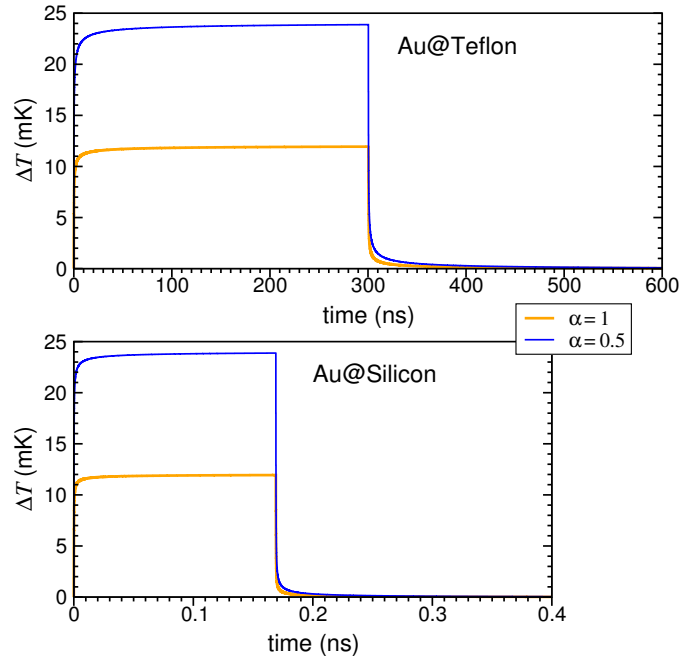}
%\smallskip
\caption{\label{fig:06}
Temperature rise ${\Delta}T$ versus heating/cooling time $t$ for a GNP of radius
$R_0=4.25$~nm in Teflon$^{\tiny\textregistered}$ and silicon spherical shells for two values
of $Q_{\rm T}$ (for Teflon$^{\tiny\textregistered}$) and $Q_{\rm Si}$ (for silicon) from
Table \ref{tab:04}. Upper curves correspond to emissivity factor of GNPs $\alpha=0.5$,
lower curves -- to $\alpha=1$.
}
\end{figure}

\begin{figure}[t!]
\includegraphics[width=0.49\textwidth]{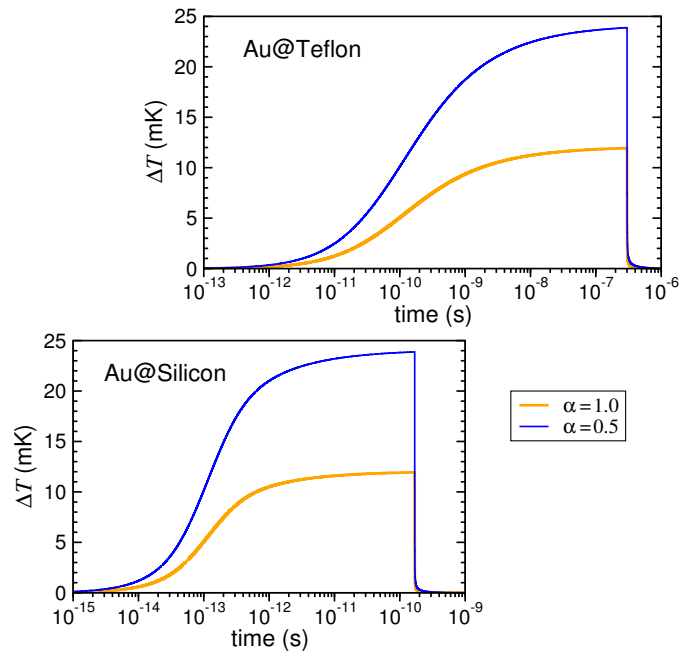}
\caption{\label{fig:07}Similar to Fig.~\ref{fig:06}, using the logarithmic time scale.}
\end{figure}

Being also interested in the detector's reaction time, we
turn now to time-dependent solution of Eq.~(\ref{eq:05}), using the ``threshold'' $Q$ values
from Table \ref{tab:04}.
The time evolution of temperature at the center of GNP (towards the target value for
the steady solution), ${\Delta}T(t,r=0)$, are shown in Fig.~\ref{fig:06}
(in the linear time scale) and in Fig.~\ref{fig:07} (in the logarithmic scale).
The heating phase was set on until the ${\Delta}T$ value reached an (arbitrarily chosen) level of 99.5\%
of the saturatton value, after which the heating was set off. The resulting ``response functions''
${\Delta}T(t)$ provide a measure of the detector's characteristic reaction time. In silicon,
thermal conductivity is much higher than that in Teflon$^{\tiny\textregistered}$ 
(see Table \ref{tab:03}). Accordingly, the heating/cooling times of 8.5~nm diameter GNPs
in silicon matrix are much shorter than those in Teflon$^{\tiny\textregistered}$ matrix.
In the logarithmic scale (Fig.~\ref{fig:07}), one can easily see that the characteristic time
at which half of the target temperature is attained is by three orders of magnitude
longer in the Teflon$^{\tiny\textregistered}$ matrix ($\sim$10$^{-10}$~s)
than in silicon ($\sim$10$^{-13}$~s). However, even for
the Teflon$^{\tiny\textregistered}$-based converter the response time
is likely to be well below microseconds and hence sufficient for real-time imaging
of tissue or patient. The ``bonus'' of the ``slow'' Teflon$^{\tiny\textregistered}$ matrix is
that much lower THz power levels are required to achieve comparable heating.

Because of its very short heating / cooling time, the THz-to-IR converter on the GNP / silicon basis
might be used for registering fast processes in the course of which an appreciable power
of THz radiation is released. For medical imaging, on the contrary,
the GNP / Teflon$^{\tiny\textregistered}$ converter seems to be advantageous, 
since it would afford an acceptable registration rate even at THz powers two orders of magnitude
lower than those needed in combination with the silicon matrix.

As a general observation based on a number of trial calculations
for a nanoparticle immersed in the matrix of as low thermal conductivity as that of 
Teflon$^{\tiny\textregistered}$,
the exact choice of transport parameters for the nanoparticle 
(confinement-dependent and estimated for a given particle size vs bulk)
becomes relatively irrelevant. As is evident from Eq.~(\ref{eq:05}),
the temperature evolution would follow the uniform scaling of the heat source term,
therefore the response time wouldn't depend on the emissivity factor
of individual particles, but only on the GNP size and the embedding material.
The temperature sensitivity of the IR camera may be a limiting factor, though.

\subsection{On enhancement of conversion efficiency of THz radiation into heat in GNPs}
\label{subsec:enhancement} % (4.4) -----------------------------------------------
We selected gold as promising metal for nanoparticles because it does not oxidise at room
temperatures, hence, surfaces of GNPs are clear of oxides, which could absorb THz radiation.
However, in gold, a number of Fermi electrons able to absorb the THz photons is not large. We
think, it is possible to enhance the number of electrons at the Fermi level of gold by two ways: 
(a) by introducing impurity atoms like Fe or Ta, which form electron $d$ states
at the Fermi level of gold \cite{JNP6-061709}; 
(b) by alloying gold with palladium, which possesses peak of electron DOS at the Fermi
level \cite{PRB14-3446} and form continuous row of solid solutions with gold \cite{BullAlloy6-229}.
Palladium, like gold, does not oxidise in air at room temperatures.
The alloys with compositions close to
$\sim\,$50~at.\%Au -- 50~at.\%Pd seem promising in the sense of being the most distorted,
because they are ``as much as possible away'' from lattices of pure constituents. 
The distortion induces intense scattering of electrons, which may absorb the energy of THz photons
and channel this energy into lattice vibrations. Consequently, the conversion of THz radiation
into heat is expected to be enhanced by alloying. Indeed, the de Broglie wavelength 
of the Fermi electrons (as well as electrons, which absorbed the THz photons)
is $\sim\,$0.5~nm. This happens to be reasonably commensurate with the expected areas
with distorted periodicity in the alloy made out of Au and Pd, characterized by
lattice parameters of 0.408~nm and 0.389~nm, respectively \cite{BullAlloy6-229}.
The following observations summarize the arguments as for the ``usefulness'' of alloying:

%\begin{enumerate}
%\item
$(i)$ as compared to pure gold, the Au-Pd alloys possess enhanced electron DOS at the Fermi level
\cite{SurfInterfAna36-793,PRB58-9817,JElecSpec78-43}; in these conditions, due to a thermal smearing
of the DOS distribution,
the electron states near to the Fermi level are occupied partially. Therefore, in Au-Pd alloys,
on the one hand, a number of Fermi electrons-absorbers of THz photons is larger than in pure gold,
and on the other hand, an enhanced number of unoccupied electron states close by the Fermi level
causes an intensive scattering of excited electrons (which absorbed THz photons);

$(ii)$ the specific electrical resistivity of the Au-Pd alloys are several times higher than that
of pure gold; moreover, at compositions $\sim\,50$~at.\%Au -- 50~at.\%Pd,
it reaches maximum values and one order of value higher than that of pure gold 
\cite{GoldBull5-74,JPhysF4-2189}.

\section{Discussion}
\label{sec:discus} % (# 5) ==========================================
When operating the schemes shown in Figs.~\ref{fig:01} and \ref{fig:02},
the previously suggested THz emitters \cite{BeilJNano7-983,Nanotechnology29-285704}
and the present THz-to-IR converter are intended to be used in tandem.
The advantage would be that the GNPs in the THz-to-IR converter would absorb THz photons 
emitted by gold nanoobjects of the emitter, \emph{i.e.} photons of the same energies
(1.57 or 36.0~meV). This absorption (by the Fermi electrons of 8.5~nm diameter GNPs)
would happen resonantly and directly, \emph{i.e.} without contributions of LAVMs considered 
in Ref.~\onlinecite{FANEM2015}.

Estimated length $\times$ width sizes of both Teflon$^{\tiny\textregistered}$ and silicon matrices
are 9.6~mm $\times$ 7.7~mm.
Thermal conductivity of silicon is much higher than that of Teflon$^{\tiny\textregistered}$,
therefore, in Table \ref{tab:04}, the powers $Q_{\rm Si}$ are much larger
than $Q_{\rm T}$, which results in shorter heating/cooling times as compared to those
in Teflon$^{\tiny\textregistered}$ matrix-based the THz-to-IR converter 
(see Figs.~\ref{fig:06} and \ref{fig:07}).

Estimated heating/cooling times of 8.5~nm diameter GNPs ($\sim\,$169~ps / 1.16~ns
and $\sim\,$300~ns / 1.25~$\mu$s in, respectively, silicon and Teflon$^{\tiny\textregistered}$ matrices)
show that the considered THz-to-IR converters would response to any object's evolution instantly,
for all practical applications envisaged. For technical operation, either a pulsed
THz radiation source (the pulse duration $\geq\,$169~ps for silicon matrix 
and $\geq\,$300~ns for Teflon$^{\tiny\textregistered}$ matrix, with pauses between
pulses longer than, correspondingly, 1.16~ns and 1.25~$\mu$s), or a continuous
THz radiation source would be required. In this context, the use of the THz radiation sources
described in Ref.~\onlinecite{BeilJNano7-983,Nanotechnology29-285704} would be a good choice.
The conventional powers required for operating the THz-to-IR converters built around
the Teflon$^{\tiny\textregistered}$ and silicon matrices are correspondingly about
110~$\mu$W and 32.3~mW, that is, within accessible limits of contemporary
THz radiation sources \cite{Proc2004FEL-216}.

Kuznetsov \emph{et al.} \cite{APL99-023501,TechnMessen78-526,PIER122-93,%
ProcSPIE8423-15,JPhysConfSer490-012174,TechPhysLett42-1130,SciRep6-21079} 
used the THz-to-IR converter with a topological pattern of split-ring resonators.
Its drawbacks are the so-called ``comet tail'' and image blooming effects,
which deteriorate the converter's response time and spatial resolution. No such effects
are likely to occur in the THz-to-IR converter based on GNPs. Indeed, in the Teflon$^{\tiny\textregistered}$ 
and silicon matrices, the edge of a reference cuboid accounted for one GNP is equal to
$$
(N^{\ast}_{\rm T})^{-1/3} = (N^{\ast}_{\rm Si})^{-1/3} = (4.44\,{\cdot}10^4\;{\rm mm}^{-3})^{-1/3}
\approx 2.8\,{\cdot}10^4\;{\rm  nm},
$$
therefore, the distance
$0.5{\cdot}(N^{\ast}_{\rm T})^{-1/3} = 0.5{\cdot}(N^{\ast}_{\rm Si})^{-1/3} = 1.4\,{\cdot}10^4\;{\rm nm}$
exceeds by far the maximum radii of radial distributions of temperature around heated GNPs
shown in Fig.~\ref{fig:06}. Hence, the temperature fields of neighboring heated GNPs would not overlap.
Moreover, the cooling times of GNPs are quite small. Therefore, no effects like the ``comet tail''
or the image blooming are \emph{a priori} expected.

Characteristics of the FPA imaging devices are being improved, new types of FPAs
are now making their appearance with temperature sensitivity 
${\Delta}T_{bb}=10$~mK \cite{DALI_D900-Series} and pixel's size $d=10$~$\mu$m \cite{Daphnis-HD-MWIR}.
With this, the power sensitivities of the THz-to-IR converters would be correspondingly enhanced.

We have patented our approaches to conversion of THz vibrations into THz
electromagnetic radiation \cite{Patent-RU2650343},
the THz-to-IR converter \cite{Patent-RU2642119},
as well as the source of THz radiation \cite{Patent-RU2622093}.
Also, we have published our papers \cite{Nanotechnology29-285704,Ferroel509-158}
related to these issues and now hope that these works would promote introducing
the THz imaging method of the human skin cancer to the modern medical practice.

\section{Conclusion}
\label{sec:conclu} % (# 6) ===========================================
Contrary to ionizing X-rays, the THz radiation is a non-ionizing one and hence not harmful to living
organisms. The THz medical imaging is expected to become a promising non-invasive technique
for monitoring the human skin's status and for early detection of pathological conditions.
In this context, the development of terahertz imaging setups emerged as one of the ``hottest'' areas
in nanotechnology-supported modalities for the cancer diagnostics. The THz-assisted diagnostics in
reflection geometry allows for non-invasive (\emph{in vivo}) THz imaging
of skin cancer's surface features. It could be performed \emph{in situ}, without time losses
on standard \emph{in vitro} histological tests, which require more time. In the reflection geometry,
one could investigate only skin's surface features and depth information, because the THz radiation
is strongly absorbed by water and does not penetrate tissue to any significant depth.
In the transmission geometry, one could study the clinically prepared tissue
samples both of inner organs and the skin.

Other areas of dermatology, where the THz imaging in reflection geometry might be advantageous,
are the following: skin burns and wound inspection through bandages; 
monitoring the treatment of skin conditions (like psoriasis), since this allows to avoid
the direct contact with the skin as \emph{e.g.} under ultrasound investigation. These advantages
are moreover supported by the fact that the THz imaging is cheaper than the magnetic resonance one.

Summarizing, the currently available offer of highly sensitive infrared cameras
allows to design schemes for the THz medical imaging of malignant tumours in biological tissue
both \emph{in vitro} and \emph{in vivo}. The schemes contain two main parts, the terahertz-to-infrared
converter and the infrared thermal imaging camera. The obtained theoretical results demonstrate
that the suggested approach can be realized with the THz-to-IR converter made of
Teflon$^{\tiny\textregistered}$ film of $\sim$0.1~nm thickness, containing gold nanoparticles,
ideally of 8.5~nm diameter. 
In order for this concept to be embodied, an elaboration of the process for making the
Teflon$^{\tiny\textregistered}$ film containing gold nanoparticles is yet required.
The principles applied when designing this THz-to-IR converter could be also useful in
development of other devices for screening goods, non-destructive monitoring quality control, etc.

% --- uncomment the following two lines to run LaTeX + BibTeX :
%\bibliographystyle{ieeetr}%phase-trans}
%\bibliography{my_publist,Alloys,BioMed,THz,GNP,Patents,Phonons}
% ----
%\end{document}

\end{document}